\documentclass[preprint]{revtex4}
\usepackage{graphicx,amsmath,amssymb}
\usepackage{bm}

\newcommand{\be}{\begin{equation}}
\newcommand{\ee}{\end{equation}}
\newcommand{\bea}{\begin{eqnarray}}
\newcommand{\eea}{\end{eqnarray}}

\begin{document}

\begin{titlepage}
\title{Coarse-grained forms for equations describing\\ the microscopic
motion of particles in a fluid}
\author{Shankar P. Das${}^1$
\email{shankar@mail.jnu.ac.in} and Akira Yoshimori${}^2$}
\affiliation{${}^1$ School of Physical Sciences, Jawaharlal Nehru
University, New Delhi 110067, India.\\ ${}^2$ Department of Physics,
Kyushu University, Fukuoka 812-8581, Japan}

\begin{abstract}

Equations of motion for the microscopic number density
$\hat{\rho}({\bf x},t)$ and the momentum density $\hat{\bf g}({\bf
x},t)$ of a fluid have been obtained in the past from  the
corresponding Langevin equations representing the dynamics of the
fluid particles. In the present work we average these exact
equations of microscopic dynamics over the local equilibrium
distribution to obtain stochastic partial differential equations for
the coarse grained densities with smooth spatial and temporal
dependence. In particular, we consider Dean's exact balance equation
for the microscopic density of a system of interacting Brownian
particles to obtain the basic equation of the dynamic density
functional theory. In the thermally averaged equation for the coarse
grained density ${\rho}({\bf x},t)$, the related dependence on the
bare interaction potential in Dean's equation is converted to that
on the corresponding direct correlation functions of the density
functional theory.

\end{abstract}

\vspace*{1cm}

\pacs{05.10.Gg,05.20.Jj,05.40.-a,47.10.ab}

\maketitle
\end{titlepage}


The equations of fluctuating nonlinear hydrodynamics (FNH)
\cite{LL,forster,gfm} provide a very useful description of the time
dependent behavior of various types of liquids, ranging from simple
liquids \cite{kirkpatrick-nieu}, solutions\cite{Fredrickson,Doem},
supercooled metastable
liquids\cite{das-mazenko,dufty-sch,yeo,kim-kawasaki},
mixtures\cite{spd-harbola}, and complex
liquids\cite{marcheti,sriram-rev}. These equations primarily
represent the time evolution of the respective coarse grained
densities for the system. Experimentally measurable correlation
functions of these densities are obtained from the equations of FNH
by averaging the corresponding quantities with respect to the noise.
The simplest example of a collective density is the number density
$\rho({\bf r},t)$ of particles in a fluid. The latter is obtained as
the average of the corresponding microscopic quantity,
$\hat{\rho}({\bf x},t){\equiv}{\rho}({\bf x},t)$, where the  hat
over $\rho$ indicates its dependence on the phase space coordinates
of the fluid particles and the angular brackets denote an averaging
over the appropriate ensemble. Dynamic equations for the microscopic
densities (with a hat) are obtained exactly from the corresponding
equations of motion of the individual fluid particles. For example,
if the fluid particles follow reversible Newtonian dynamics (ND)
the  corresponding  exact balance equations of hydrodynamics are
known as Euler's equations which are also time reversible. For fluid
particles with dissipative dynamics, the corresponding equations for
the microscopic densities \cite{dean,yoshimori} are also time
irreversible. Apart from being exact representations of the
corresponding microscopic dynamics, these exact equations are formal
and needs to be averaged over an appropriate ensemble for practical
use. In this paper we obtain coarse grained forms of these exact
equations by averaging them over a suitable local equilibrium
distribution. The final equations are stochastic partial
differential equations having smooth spatial and temporal
dependence. The coefficients involved in the FNH equations are
related to the thermodynamic properties of the fluid.



We begin by considering a $N$ particle system in which the
microscopic dynamics of the constituent particles is described by
Smoluchowski equation \cite{smoluchowski} involving only the
particle coordinates. Let the $\alpha$-th   particle (for
$\alpha=1,...N$)  be of mass $m$ and its position coordinates are
given by $\{ {\bf x}_\alpha \}$. A colloidal system with heavy
particles in a solution is a typical example for such a system. In
the over damped limit, the time dependence of the momentum of the
particle is ignored and the stochastic equation of motion for the
coordinate ${\bf x}_\alpha$ of the $\alpha$-th particle is obtained
as,

\begin{equation}\label{BD-leqn} \gamma_0 \frac{d{\bf x}_\alpha(t)}{dt}
=  - \sum_{\nu=1}^N {\bf\nabla}_\alpha U({\bf x}_\alpha-{\bf x}_\nu)+
\bm{\xi}_\alpha(t).\end{equation}

\noindent The quantity $\gamma_0$ which has dimension of
$[\mathrm{M}\mathrm{T}^{-1}]$ is chosen to be unity in the
following. $U({\bf x}_\alpha-{\bf x}_\nu)$ is the interaction
potential between particles $\alpha$ and $\nu$. The symbol
$\nabla_\alpha$ with the Greek subscript $\alpha$ denotes the
derivative operator with respect to the corresponding component of
${\bf x}_\alpha$. The stochastic part or the noise ${\bm
\xi}_\alpha(t)$ in the equation of motion (\ref{BD-leqn}) is white
and gaussian. The noise correlation is given by,

\begin{equation}\label{mnoise} <\xi_\alpha^i(t) \xi_\beta^j(t^\prime)>
= 2k_BT \delta_{\alpha\beta}\delta_{ij}\delta (t-t')~~. \end{equation}

\noindent The collective density $\hat{\rho}({\bf x},t)$ is defined
in terms of phase space variables $\{{\bf
x}_\alpha\}{\equiv}{\Gamma_N}(t)$ as:

\be \label{def-den} \hat{\rho}({\bf x},t) = \sum_{\alpha=1}^{N}
\delta ({\bf x}-{\bf x}_\alpha(t))~~. \ee

\noindent Since the number $N$ of particles is constant
$\hat{\rho}({\bf x},t)$ is a microscopically conserved quantity.
Dean\cite{dean,yoshimori1} obtained the following equation of motion
for $\hat{\rho}$ :

\be \label{Dn-eq1} \frac{\partial}{\partial{t}}\hat{\rho}({\bf x},t)
= k_BT \nabla^2 \hat{\rho}({\bf x},t) + {\bf \nabla} \cdot
\{\hat{\rho} ({\bf x},t) {\bf \nabla} \int d{\bf x}'U({\bf x}-{\bf
x}') \hat{\rho}({\bf x}',t)\} + \hat{\zeta} ({\bf x},t)~~, \ee

\noindent where ${\bf \nabla}$ denotes the derivative operator with
respect to components of ${\bf x}$. The correlation of the random
force denoted by $\hat{\zeta}({\bf x},t)$ is obtained as,

\begin{equation}\label{BD-ns1} \hat{\zeta} ({\bf x},t) =
-\sum_\alpha {\bf \nabla}_\alpha \cdot \{ \delta ({\bf x}-{\bf
x}_\alpha(t))\bm{\xi} _\alpha(t)\} = {\bf \nabla} \cdot \sum_\alpha
\delta ({\bf x}-{\bf x}_\alpha(t))
\bm{\xi}_\alpha(t)~~.\end{equation}

\noindent $\hat{\zeta}({\bf x},t)$ is dependent on the phase space
coordinates through ${\bf x}_\alpha$. Eqn. (\ref{Dn-eq1}) is an
exact representation of the stochastic dynamics described by eqns.
(\ref{BD-leqn})-(\ref{mnoise}). Unlike the time reversible Euler
equations, the Dean equation is dissipative since the corresponding
microscopic dynamics described by eqn. (\ref{BD-leqn}) is
dissipative. We average eqn. (\ref{Dn-eq1}) to obtain a partial
differential equation for the corresponding coarse grained density
$\rho({\bf x},t)$  which has smooth spatial and temporal dependence.
This stochastic equation describes the time dependence of density
fluctuations in a non-equilibrium state. The averaging should
therefore be done over the corresponding non-equilibrium ensemble.
We approximate the latter in terms of the local equilibrium
ensemble. This is motivated from the fact that the hydrodynamic
description of the non-equilibrium fluid corresponds to the time
regime in which it has reached a state of local equilibrium. At this
stage the local densities  are sufficient to describe the state of
the system. The average of the microscopic densities (dependent on
phase space coordinates) over the local equilibrium distribution
function $f_\mathrm{l.e}$ obtains the corresponding hydrodynamic
field with smooth spatial and temporal dependence.

We note that the temperature $T$ enters the exact eqn.
(\ref{Dn-eq1}) only through the first term on the RHS. This is a
consequence of using the chain rule of the stochastic differential
equations of the I\^{t}o calculus in obtaining eqn. (\ref{Dn-eq1}).
Let a set of stochastic variables $x_i(t)$ ($i=1,..,m$) satisfy the
stochastic equation $\dot{x}_i=h_i +g_{ij}\xi_j$ with the
correlation of the white noise $\xi_i$ being given by
$<\xi_l(t)\xi_m(t')>=\delta_{lm}\delta(t-t')$. I\^{t}o chain
rule\cite{oskendal} obtains the stochastic differential equation for
the variable $y(\{x_i\})$ in the form

\begin{equation}\label{Ito-cal3} \dot{y} = \sum_i
\frac{\partial{y}}{\partial{x_i}}\dot{x}_i + \sum_{i,j,k}
\frac{1}{2}\frac{\partial^2{y}}{\partial{x_i}\partial{x_j}}g_{ik}g_{kj}.
\end{equation}

\noindent With eqn. (\ref{BD-leqn}) for dynamics of ${\bf
x}_\alpha$, the matrix $g_{ij}=\sqrt{2k_BT} \delta_{ij}$ is
diagonal. The first term on the RHS of eqn. (\ref{Dn-eq1}) follows
from the $g_{ij}$'s in the second term on the RHS of eqn.
(\ref{Ito-cal3}). This holds even for a system of noninteracting
particles. We define next the averaging procedure with respect to
the local equilibrium distribution $f_\mathrm{l.e}$. The latter is
considered to have attained a stage when the system is at
temperature $T$. This thermalization process is closely linked to
the momentum distribution becoming the Maxwellian. So far the
momentum current density has not been included in the formulation.
The momentum current density $\hat{\bf g}$ is the flux for number
density $\hat{\rho}({\bf x},t)$ in the corresponding balance
equation representing number conservation.

\be \label{cur-def}  \frac{\partial}{\partial{t}}\hat{\rho}({\bf
x},t)= - {\bf \nabla} . \sum_\alpha {\bf p}_\alpha \delta({\bf
x}-{\bf r}_\alpha(t)) = - {\bf \nabla} . \hat{\bf g}({\bf x},t) \ee

\noindent The probability function for the local equilibrium state
is obtained in analogy with that of the equilibrium state
\cite{morozov}. We consider here the distribution function
$f_\mathrm{l.e}$ in terms of the corresponding local thermodynamic
variables $\{\beta({\bf x},t),\mu({\bf x},t),v({\bf x},t)\}$. The
distribution $f_\mathrm{l.e} (\Gamma_N^\prime;t)$ in the local rest
frame, moving with the local velocity ${\bf v}({\bf x},t)$, is
obtained as
\begin{equation}\label{le-dprime1} f_\mathrm{l.e} (\Gamma_N^\prime;t)
= Q^{-1} {\exp}\Big ( - \int d{\bf r} \beta({\bf r},t)
\Big [\hat{e}^\prime ({\bf r}) - \mu ({\bf r},t) \hat{\rho}^\prime
({\bf r}) \Big ] \Big). \end{equation}

\noindent where $Q$ is the partition function $Q=\int d\Gamma
{f_\mathrm{le}}(\Gamma_N,t)$. The corresponding microscopically
conserved densities of the energy and number of particles in the
local frame are respectively denoted by $\hat{e}^\prime$ and
$\hat{\rho}^\prime$. The fluid has attained a state with a fixed
temperature $T$ ($\beta({\bf r},t){\equiv}\beta=1/(k_BT))$ and the
corresponding local equilibrium distribution is given by
$f_\mathrm{l.e}(\Gamma_N^\prime;t) = Q^{-1} {\exp}\Big ( - \beta
\Big [\mathrm{H^\prime} - \int d{\bf r} \mu ({\bf r},t)
\hat{\rho}^\prime ({\bf r}) \Big ] \Big )$. The equal time
correlation of the momentum density ${\bf g}^\prime({\bf x},t)$ in
the local rest frame is now obtained as,

\be \label{gg-corln} {\langle \hat{g}^\prime_i({\bf
x}_1,t)\hat{g}^\prime_j({\bf x}_2,t) \rangle }_\mathrm{l.e}
=\delta({\bf x}_1-{\bf x}_2)\delta_{ij} {\rho}({\bf x}_1,t) k_BT \ee

\noindent where  $\rho({\bf x},t){\equiv}{< \hat{\rho}({\bf
x},t)>}_\mathrm{l.e}$ is the local density.

Having defined the distribution function we now average the
microscopic eqn. (\ref{Dn-eq1}) over the local equilibrium ensemble
to obtain an equation for the coarse grained particle density in the
following form: \be \label{BD-ceqn} \frac{\partial{\rho}({\bf
x},t)}{\partial{t}} = k_BT \nabla^2 {\rho}({\bf x},t) + {\bf \nabla}
\cdot \Big [ \int d{\bf x}' \{ {\bf \nabla} U({\bf x}-{\bf x}')\}
{<\hat{\rho}({\bf x},t) \hat{\rho}({\bf x}',t)>}_\mathrm{l.e} \Big ]
+ \zeta({\bf x},t)~~. \ee \noindent The thermally-averaged equation
(\ref{BD-ceqn}) has stochastic and regular parts. The stochastic
part or the noise $\zeta({\bf x},t)$ is obtained by coarse graining
of the microscopic quantity $\hat{\zeta}({\bf x},t)$ defined in
eqn.(\ref{BD-ns1}). Correlation of $\zeta$ is understood as a
combination of two steps. First, in $\hat{\zeta}$ we average the
noise $\xi$ of the microscopic equations of motion over the
different configurations of the Brownian particles so as to obtain
the coarse grained noise $\zeta$. This is indicated with a subscript
$\mathrm{l.e}$ on the angular brackets. Second, we correlate the
coarse grained noise $\zeta$ at two different space time points
while the equilibrium temperature of the bath is maintained at $T$.
We denote the latter with a subscript $B$ on the angular brackets to
indicate the averaging over the bath variables. The correlation of
the noise $\zeta$ in the coarse grained equation is obtained by
interchanging the order of the two operations stated above,

\begin{eqnarray} \label{cnoise-demo} \left\langle \zeta({\bf x},t)
\zeta({\bf x'},t')\right\rangle &=& {\left\langle {\left \langle
\hat{\zeta}({\bf x},t)\right \rangle}_\mathrm{l.e} {\left\langle
\hat{\zeta}({\bf x'},t') \right \rangle}_\mathrm{l.e}
\right\rangle}_\mathrm{B} \nonumber \\&\approx& {\left\langle
{\left \langle \hat{\zeta}({\bf x},t)\hat{\zeta}({\bf x'},t')
\right \rangle}_\mathrm{l.e}\right\rangle}_\mathrm{B} \nonumber
\\&=&2k_{B}T \nabla \rho({\bf x},t)
\nabla^\prime\delta({\bf x}-{\bf x'})\delta (t-t')~~.
\end{eqnarray}

\noindent $\nabla^\prime$ denotes the derivative operator with
respect to the components of ${\bf x}^\prime$. In the Markovian
approximation of large separation between the characteristic time
scales of the solute and that of the solvent variables, the noise
correlation is independent of the coarse graining process.


The first and second terms in the RHS of eqn. (\ref{BD-ceqn}) are in
terms of slowly varying quantities. These are conveniently expressed
in terms of the Liouville operator ${\cal L}$. For the $N$ particle
system ${\cal L}$ is defined as \be \label{liou-def}  i{\cal L} =
\sum_{\alpha=1}^N \left [ \frac{\partial \mathrm{H}} {\partial {\bf
p}_\alpha} \frac{\partial}{\partial {\bf r}_\alpha} -\frac{\partial
\mathrm{H}}{\partial {\bf r}_\alpha}\frac{\partial}{\partial {\bf
p}_\alpha} \right ]~~, \ee

\noindent where the Hamiltonian $\mathrm{H}=\mathrm{K}+\mathrm{V}$
is a sum of kinetic and potential parts. The momentum dependent or
the kinetic energy part is $\mathrm{K}=\sum_{\alpha=1}^N {\bf
p}_\alpha^2/(2m)$ and the interaction part $\mathrm{V}$ is a
function of the particle coordinates only. The time rate of change
of $\hat{\rho}({\bf x},t)$ is expressed in terms of the operator
${\cal L}$ to obtain $i{\cal L}\hat{\rho}=-{\bf \nabla}.\hat{\bf
g}=\dot{\hat{\rho}}({\bf x},t)$. Note however that if ${\cal L}$
acts on the momentum current density $\hat{\bf g}$, it does not
obtain the corresponding time derivative. Using the definitions
(\ref{liou-def}) and (\ref{cur-def}) respectively of the Liouville
operator ${\cal L}$ and the current ${\bf g}^\prime$ in the local
rest frame, we obtain

\be i{\cal L} \hat{g}_i^\prime({\bf x}) = -\sum_j\nabla_j
\sum_{\alpha=1}^{N} {p'}^i_\alpha {p'}^j_\alpha \delta ({\bf x}-{\bf
x}_\alpha) + \int d {\bf x}' \{ {\nabla}_i U({\bf x}-{\bf x}')\}
{\hat{\rho}({\bf x},t)\hat{\rho}({\bf x}',t)}~~. \ee

\noindent Next taking an average over the local equilibrium ensemble
in the co-moving frame, we obtain

\bea \label{two-terms}  {\langle -i{\cal L} \hat{g}^\prime_i({\bf
x},t)\rangle}_\mathrm{l.e} &=& {k_BT}\nabla_i \rho({\bf x},t) + \int
d {\bf x}' \{ \nabla_i U({\bf x}-{\bf x}')\} {<\hat{\rho}({\bf
x},t)\hat{\rho}({\bf x}',t)>}_\mathrm{l.e} ~~.\eea

\noindent Since ${\cal L}$ involves derivative operators,
integrating by parts and using the property $i{\cal L}\mathrm{H}=0$,
we obtain

\bea  {\langle -i{\cal L} \hat{g}^\prime_i({\bf x})
\rangle}_\mathrm{l.e} &=& \int d\Gamma \hat{g}^\prime_i({\bf x})
\frac{1}{Q} i{\cal L} {\exp}\Big [ -\beta \{ \mathrm{H}-\int d{\bf
x}^\prime \mu ({\bf x}^\prime)\hat{\rho}({\bf x}^\prime) \} \Big ]
\nonumber \\ &=& \beta\int d{\bf x}^\prime  \mu ({\bf x}^\prime)
{\langle \hat{g}^\prime_i({\bf x})i{\cal L}\hat{\rho} ({\bf
x}^\prime)\rangle }_\mathrm{l.e}\nonumber \\\label{liou-g} &=&
-\beta\int d{\bf x}^\prime  \mu ({\bf x}^\prime)\sum_j {\langle
\hat{g}^\prime_i({\bf x}) {\nabla}_j^\prime \hat{g}^\prime_j ({\bf
x}^\prime) \rangle }_\mathrm{l.e} = \rho({\bf x})\nabla_i \mu({\bf
x}) ~~.\eea

\noindent In reaching the last equality we have used the key
relation (\ref{gg-corln}) for momentum correlation in the local
equilibrium distribution which has thermalized at temperature $T$.

The quantity on the RHS of eqn. (\ref{liou-g}) is a thermodynamic
property. We link this to the Helmholtz free energy $F$ expressed as
a functional of the inhomogeneous density $\rho({\bf x})$. Using the
equilibrium relation $F-G=\Omega\equiv{-PV}$, where $\Omega$ is the
thermodynamic potential, we have in the density functional formalism
\cite{tvr,spd-book}

\begin{equation} F[\rho({\bf x})] \equiv \Omega [\rho({\bf x})] +\int d{\bf x}\rho({\bf x})\mu({\bf x})~~. \end{equation}

\noindent  $\Omega[\rho]$ is a functional of the density obtained
from the equivalent result of grand canonical ensemble partition
function $\Omega [\rho({\bf r})] \equiv -k_B T\ln \Xi$. The density
functional theory identifies the equilibrium density by minimizing
the grand potential $[\delta\Omega/\delta\rho({\bf x})]=0$. Using
the above relations it then follows that the corresponding Helmholtz
free energy functional satisfies $[\delta{F}[\rho]/\delta\rho({\bf
x})]=\mu({\bf x})$. Eqn. (\ref{BD-ceqn}) then reduces with the help
of eqns. (\ref{two-terms}) and (\ref{liou-g}) to the form \be
\label{BD-reqn} \frac{\partial{\rho}}{\partial{t}} = D_0{\bf \nabla
} \cdot \Bigg [ \rho {\bf \nabla}\frac{\delta{F}}{\delta\rho} \Bigg
] + \zeta~~.\ee \noindent We have now put the bare diffusion
constant $D_0=\gamma_0/(\beta{m})$ in the RHS above to keep the
dimensional factor explicit\cite{ddft,munakata}. The free energy
$F[\rho]$ of the density functional theory is expressed ( in units
of $k_BT$) as a sum of two parts
$F[\rho]=F_\mathrm{id}+F_\mathrm{ex}$, respectively denoting the
non-interacting or ideal gas part $F_\mathrm{id}[\rho]$ and the
interaction part $F_\mathrm{ex}[\rho]$. The ideal gas part is
obtained as

\be \label{F_ideal}
 F_\mathrm{id}[\rho] = k_BT \int d{\bf x}
 \rho({\bf x})[\ln(\Lambda_0^3 \rho({\bf x})) -1]~~.
\ee

\noindent where $\Lambda_0=h/\sqrt{2\pi{m}k_BT}$ is the thermal wave
length. The interaction part or the so called excess part
$F_\mathrm{ex}[\rho]$ is generally expressed in a functional Taylor
expansion in terms of direct correlation functions $c^{(i)}({\bf
x}_1,...,{\bf x}_i)$ at a density $\rho_0$. \be \label{dirc-c-defn}
c^{(i)}({\bf x}_1, ...,{\bf x}_i;\rho_0)=\left [\frac{\delta^i
F_\mathrm{ex}[\rho]}{\delta \rho({\bf x}_1)...\delta \rho({\bf
x}_i)} \right ]_{\rho=\rho_0} \ee

\noindent The two point function for $i=2$ is the Ornstein-Zernike
direct correlation function $c^{(2)}$.

Going beyond the over damped limit considered above, the equations
of motion of the fluid particles are obtained in terms of both the
respective momentum and position coordinates. This is termed as the
Fokker-Planck dynamics (FPD). For the $N$ particle system the
momentum ${\bf p}_\alpha$ of the $\alpha$-th particle is taken as
$\dot{\bf x}_\alpha(t)$ ( mass taken as unity) and its time rate of
change is

\be \label{FPD-leqn} \frac{d{\bf p}_\alpha(t)}{dt} = -
\sum_{\beta=1}^N {\bf \nabla}_\alpha U({\bf x}_\alpha(t)-{\bf
x}_\beta(t))-{\gamma}_0 {\bf p}_\alpha + \bm{\xi}_\alpha(t), \ee

\noindent where ${\gamma}_0$ is a dissipative coefficient. Note that
this model does not conserve momentum microscopically unless the
dissipative coefficient ${\gamma}_0$ has a $\nabla^2$ operator
associated with it. Balance equations for the collective densities
$\{\hat{\rho}({\bf x},t),\hat{g}({\bf x},t)\}$ have been obtained by
Nakamura and Yoshimori\cite{yoshimori}.  By averaging these
microscopic balance equations, we obtain the corresponding equations
of motion for the coarse grained variables $\{\rho({\bf x},t),{\bf
g}({\bf x},t)\}$.

\begin{eqnarray}
&&\label{FPD-cr}\frac{\partial}{\partial{t}}{\rho}({\bf x},t)
+ {\bf \nabla}\cdot{\bf g}({\bf x},t) = 0  \\
\label{FPD-colc2} &&\frac{\partial}{\partial{t}} {g}_i({\bf x},t) +
{\gamma}_0 {g}_i({\bf x},t) + {\nabla}^j{\left \langle \{
\sum_\alpha \frac{p^i_\alpha p^j_\alpha}{m}  \delta ({\bf x}-{\bf
x}_\alpha(t)) \}\right \rangle}_\mathrm{l.e} \nonumber \\ &+& \int
dx' \{ {\nabla}_i U({\bf x}-{\bf x}')\} {<\hat{\rho}({\bf
x},t)\hat{\rho}({\bf x}',t)>}_\mathrm{l.e} = {\theta}_i ({\bf
x},t)~~. \end{eqnarray} \noindent The noise $\theta_i({\bf x},t)$ in
the generalized Langevin equation is obtained by coarse graining of
the noise $\hat{\theta}_i({\bf x},t)$ which is  defined in terms of
the noise $\xi_\alpha$ in the micro-dynamic equations.
\begin{equation}
\label{ndefn1} \hat{\theta}_i ({\bf x},t)= \sum_{\alpha=1}^{N}
\delta ({\bf x}-{\bf x}_\alpha(t))\xi^i_\alpha (t)~~. \end{equation}

\noindent Using the same approximations as discussed with eqn.
(\ref{cnoise-demo}), the correlation of the multiplicative noise
$\theta_i({\bf x},t)$ is obtained as, \be \label{col-ncora}
\left\langle  \theta_i({\bf x},t)\theta_j({\bf x'},t')\right\rangle
= 2k_{B}T {\gamma}_0 {\rho}({\bf x},t) \delta({\bf x}-{\bf
x'})\delta (t-t')~~. \ee \noindent We evaluate the third term on the
LHS of eqn. (\ref{FPD-colc2}) by making a change of variables ${\bf
p}_\alpha = {\bf p}^\prime_\alpha +m {\bf v} ({\bf
r^\prime}_\alpha)$ and ${\bf r}_\alpha = {\bf r}^\prime_\alpha$ in
the locally moving frame with velocity ${\bf v}({\bf
r^\prime}_\alpha)$. Using the symmetry of the distribution function
(\ref{le-dprime1}) for the locally moving frame under the ${\bf
p}^\prime {\rightarrow}-{\bf p}^\prime$ transformation this term
reduces to \be \label{FPD-ac3}  k_BT \nabla_i \rho ({\bf x},t)
+\nabla_j \left [ \frac{g_i({\bf x},t) g_j({\bf x},t)}{\rho({\bf
x},t)} \right ] ~~.\ee

\noindent Substituting the above result in eqn. (\ref{FPD-colc2}),
and using the relations (\ref{two-terms})-(\ref{liou-g}), which also
hold for the Fokker-Planck dynamics, we obtain the following
stochastic equation for the coarse grained momentum density \be
\label{FPD-cg} \frac{\partial{g_i}}{\partial{t}} + \nabla_j \left [
\frac{g_i g_j}{\rho} \right ] + {\rho}{\bf \nabla }_i
\frac{\delta{F}}{\delta\rho} + {\gamma}_0 {g}_i= {\theta}_i \ee

\noindent Eqn. (\ref{FPD-cg}) is very similar in form to the
corresponding equations of fluctuating nonlinear hydrodynamics for
the particles which follow reversible
ND\cite{das-mazenko,kawasaki-miyazima}. There is however one
important difference. Unlike in the case of the FPD, for the ND the
dissipative term involves a $1/\rho$ nonlinearity
\cite{das-mazenko}. On the other hand, in the FPD the noise
correlation given by eqn. (\ref{col-ncora}) involves the density and
hence the noise is multiplicative. The presence  of the $1/\rho$
nonlinearity in the ND case is important with respect to the
ergodicity-nonergodicity (ENE) transition of the self-consistent
mode coupling theory \cite{rmp,reichman} of dense liquids. The
latter has been extensively used for studying glassy relaxations
\cite{biroli-epl,szamel}. The ENE transition being referred to is a
consequence of feedback effects from slowly decaying density
fluctuations in a dense liquid. It has been established
\cite{das-mazenko} that the $1/\rho$ nonlinearity in the generalized
Langevin equation is essential in smoothing off this transition. For
Newtonian dynamics of particles, the single equation (\ref{BD-reqn})
for $\rho$ has also been obtained\cite{kawasaki-miyazima} by
eliminating the current ${\bf g}$ from a field theoretic formulation
of the problem. This however requires in the ND case applying the so
called adiabatic approximation in which momentum fluctuations decay
much faster than the density fluctuations.


In summary, the present work fills up an important gap between a) a
nonlinear diffusion equation for the coarse grained density field in
terms of thermodynamic direct correlation functions and b) the exact
representation of the Brownian dynamics of a system of particles
(e.g.   Dean's equation). Thus the single equation (\ref{BD-reqn})
for $\rho({\bf x},t)$  is the coarse grained form of Dean's exact
equation for the microscopic density $\hat{\rho}({\bf x},t)$. The
first equation has been used in a large number of works on fluids
and is well known to the community as an ingredient of the dynamic
density functional theory. The driving free energy for this coarse
grained equation is obtained here in the standard form of density
functional theory. The role of the bare interaction potential
${\beta}U(r)$  in Dean's equation is replaced by the corresponding
direct correlation function $c(r)$. The averaged equation which
forms the basis for field theoretic models for the dynamics of
fluids, has often been considered (without proof) to be exact for
interacting Brownian particles. The effects of nonlinearities in the
coarse grained equation are studied with various analytic techniques
which are available for calculating renormalized time correlation
functions in terms of appropriate mode coupling contributions
\cite{biroli-epl,szamel}. These models often form the basis for
studying dynamics heterogeneities and growing dynamic correlation
lengths in supercooled liquids.

\section*{Acknowledgements}  AY acknowledges the support by
Grants-in-Aid for Innovative Scientific Research Area and for
Scientific research C from the Ministry of Education, Culture,
Sports, Science and Technology of Japan. SPD acknowledges Project
2011/37P/47/BRNS of DAE for financial support. \citation{LL}


\begin{thebibliography}{99}

\bibitem{LL} L.D. Landau, and E.M. Lifshitz, Fluid Mechanics,
Oxford: Butterworth-Heinemann, 1987.

\bibitem{forster}D. Forster, {\it Hydrodynamic Fluctuations,
Broken Symmetry and Correlation Functions}, Benjamin, Reading,
Massachusetts, 1975.

\bibitem{gfm}G.F. Mazenko, {\em Nonequilibrium statistical mechanics},
Wiley-VCH New York, 2006.

\bibitem{kirkpatrick-nieu}T.R. Kirkpatrick, and J.C.~Nieuwoudt,
Phys.~Rev.~A {\bf 33}, 2651 (1986).

\bibitem{Fredrickson}G.H. Fredrickson, and E. Helfand,
J. Chem. Phys.  {\bf 93}, 2048 (1990).

\bibitem{Doem}M.W. Deem, Phys. Rev. E {\bf 51}, 4319 (1995).


\bibitem{kim-kawasaki}
B. Kim, and K. Kawasaki, J. of Stat. Mech.,  P02004 (2008).


\bibitem{das-mazenko} S.P. Das,
and G.F. Mazenko, Phys. Rev. A {\bf 34}, 2265 (1986); Phys. Rev. E
{\bf 79}, 021504 (2009).

\bibitem{dufty-sch}R. Schmitz, J. W. Dufty, and P. De, Phys. Rev.
Lett. {\bf 71}, 2069 (1993).


\bibitem{yeo}J. Yeo, Phys. Rev. E {\bf 80}, 051501 (2010).


\bibitem{spd-harbola}U. Harbola, and S. P. Das, Phys. Rev. E
{\bf 65}, 36138 (2002).

\bibitem{marcheti}T.B. Liverpool,
M.C. Marchetti, Phys. Rev. Lett. {\bf 97}, 268101 (2006).

\bibitem{sriram-rev}S. Ramaswamy, Annu. Rev. Condens. Matter Phys.,
{\bf 1}, 9.1–9.23 (2010).

\bibitem{dean} D.S. Dean, J Phys. A : Math. Gen. {\bf 29}, L613 (1996).

\bibitem{yoshimori}T. Nakamura, and A. Yoshimori, J. Phys. A : Math Theor.
 {\bf 42}, 065001 (2009).

\bibitem{smoluchowski}M. von Smoluchowski,
Annalen der Physik {\bf 48}, 1103 (1915).


\bibitem{yoshimori1} A. Yoshimori, A., Phys. Rev. E {\bf 59}, 6535 (1999);
 Phys. Rev. E {\bf 71}, 031203 (2005).

\bibitem{oskendal}
B. Oksendal, {\em Stochastic Differential equations}, Springer,
Berlin (1992).

\bibitem{morozov}D.N. Zubarev, V. Morozov and G. R\"opke,
{\em Statistical Mechanics of Non-Equilibrium Processes}, Vol. II,
Akademie Verlag Berlin, 1997.

\bibitem{spd-book}S.P. Das, {\it Statistical Physics of Liquids at Freezing and
Beyond}, Cambridge University Press, New York, 2011.


\bibitem{tvr}T.V. Ramakrishnan, and M.~Yussouff,
Phys.~Rev.~B {\bf 19}, 2775 (1979).


\bibitem{ddft}
U.M.B. Marconi, and P. Tarazona, J. Phys.: Condens. Matter 12 A41
(2000).

\bibitem{munakata}T. Munakata, Phys. Rev. E 67, 022101 (2003).

\bibitem{kawasaki-miyazima}K. Kawasaki, and S. Miyazima, Z. Phys. B,
Condensed Matter, {\bf 103}, 423 (1997).


\bibitem{rmp}S.P. Das, Rev. of Mod. Phys. Rev. {\bf 76},
785 (2004).


\bibitem{reichman}
D. R. Reichman and P. Charbonneau, J. Stat. Mech., P05013 (2005).


\bibitem{biroli-epl}
 G. Biroli and J-P Bouchaud, Europhys. Lett. 67 21 (2004).


\bibitem{szamel}G. Szamel, Phys. Rev. Lett. 90 228301 (2003).


\end{thebibliography}
\end{document}